\documentclass[preprints,review,accept,pdftex,moreauthors]{Definitions/mdpi} 
\firstpage{1} 
\pubvolume{1}
\issuenum{1}
\articlenumber{0}
\pubyear{2023}
\copyrightyear{2023}
\datereceived{13 July 2023} 
\dateaccepted{25 August 2023} 
\datepublished{} 
\hreflink{https://doi.org/} 
\pdfoutput=1

\Title{The Structure and Evolution of Stars: Introductory Remarks}

\TitleCitation{Stellar Structure}

\Author{Dominic M. Bowman$^{1,2,\ddagger, *}$\orcidA{}, Jennifer van Saders $^{3,\ddagger, *}$\orcidB{}, and Jorick S. Vink$^{4,\ddagger, *}$\orcidC{}}

\AuthorNames{Dominic M. Bowman, Jennifer Van Saders, and Jorick S. Vink}

\AuthorCitation{Bowman, D.~M.; van Saders, J.; Vink, J.S.}

\address{%
$^{1}$ \quad Institute of Astronomy, KU Leuven, Celestijnenlaan 200D, B-3001 Leuven, Belgium; dominic.bowman@kuleuven.be\\
$^{2}$ \quad School of Mathematics, Statistics and Physics, Newcastle University, Newcastle upon Tyne, NE1 7RU, UK \\
$^{3}$ \quad Institute for Astronomy, University of Hawai'i, Honolulu, HI 96822, USA; jlvs@hawaii.edu\\
$^{4}$ \quad Armagh Observatory and Planetarium, College Hill, BT61 9DG, Armagh, Northern Ireland; jorick.vink@armagh.ac.uk\\
}

\corres{Correspondence: dominic.bowman@kuleuven.be (DMB); jlvs@hawaii.edu (JvS); jorick.vink@armagh.ac.uk (JV)}

\secondnote{These authors contributed equally to this work.}

\abstract{In this introductory chapter of the Special Issue entitled 'The Structure and Evolution of Stars', we highlight the recent major progress made in our understanding in the physics that governs stellar interiors. In so doing, we combine insight from observations, 1D evolutionary modelling and 2+3D rotating (magneto)hydrodynamical simulations. Therefore, a complete and compelling picture of the necessary ingredients in state-of-the-art stellar structure theory and areas in which improvements still need to be made are contextualised. Additionally, the over-arching perspective that links all the themes of subsequent chapters is presented.}

\keyword{stellar structure and evolution; convection; rotation; diffusion; magnetism; asteroseismology; radiation pressure}

\begin{document}

\setcounter{secnumdepth}{0}
\section{Introduction}
\label{section: intro}

Our understanding of stellar structure and evolution has progressed dramatically in recent decades thanks to high-precision observations from modern instrumentation, increased complexity and flexibility in evolutionary modelling software tools, and improved physics and computing resources for advanced (magneto)hydrodynamical simulations. Such a combined approach of tackling unresolved issues from these three sides has allowed improved physical prescriptions in models calibrated by observations. On the other hand, it has also unveiled large theoretical uncertainties for stars across the Hertzsprung--Russell (HR) diagram \citep{Maeder2000a, Maeder_rotation_BOOK, Langer2012}, which is subsequently driving new advances.

While many textbooks, special volumes, and conference proceedings have been written about stellar evolution, we feel that there is a need to specifically address the physical ingredients that determine the structure of a star. Moreover, we wish to identify common themes in stellar structure across the HR~diagram, since focus is often given to a niche part of it. Even before we are able to predict the outcomes of stellar evolution as compact white dwarfs, neutron stars, or black holes, with or without a prior supernova explosion, the physics of stellar evolution up to these final stages is uncertain due to inaccurate 1D stellar structures that percolate into evolutionary predictions. 

Generally speaking, stars consist of three regimes: a core, an envelope, and an atmosphere from which the light emerges. Depending on the stellar mass and the evolutionary stage, cores and envelopes can be either radiative or convective. These regions define the (dominant) form of energy transport, but their physical definition and the interface between them represent a large source of uncertainty in stellar structure theory. Whilst stellar atmospheres are key messengers of astronomical information, they are also physical laboratories of radiation pressure leading to radiation-driven winds for high-mass stars and chemical mixing and transport phenomena such as radiative levitation in hot low-mass stars, which is where heavy elements with large cross-sections can gain momentum by absorbing photons from outflowing radiation. In this Special Issue, we focus on the basic interior structure of cores and envelopes, leaving stellar atmospheres, stellar winds, as well as nucleosynthesis to other outlets.

In Autumn of 2021, the authors of this chapter were fortunate to be present in person at the 'Probes of Transport in Stars' scientific programme organised by Matteo Cantiello, Adam Jermyn, Daniel Lecoanet, and Jamie Tayar, and hosted by the Kavli Institute for Theoretical Physics (KITP), University of California Santa Barbara (UCSB), USA. Such a fantastic opportunity to bring together a diverse collection of observers, modellers and numericists allowed new avenues of research to be conceived pertaining to various transport mechanisms within stellar structure and evolution theory. Born out this innovative and productive program, we conceived of this Special Issue to document the impactful new and ongoing projects that are at the forefront of astronomy research in the field of stellar structure theory. The chapters within this Special Issue span a range of topics related to the physics of stellar structure, with several of the authors also being in-person or virtual participants of the `Probes of Transport in Stars' program hosted by KITP in 2021.

In this manuscript, we provide a brief overview of each chapter in this Special Issue. Yet it is also our goal to link and discuss the overarching themes across all of them for the non-expert.
The order of the Special Issue is the following:

\begin{enumerate}
    \item{Joyce \& Tayar: 1D convection in stellar modelling}
    \item{Alecian \& Deal: Opacities and atomic diffusion}
    \item{Keszthelyi: Magnetism in high-mass stars}
    \item{Lecoanet \& Edelmann: Multi-D simulations of core convection}
    \item{Anders \& Pedersen: Convective boundary mixing in main-sequence stars}
    \item{Jiang: Radiation dominated envelopes of massive stars}
\end{enumerate}

In the subsequent sections, we provide a brief overview of their goals and contents as an introduction for the non-expert reader.


\section{Chapter 1: 1D convection in stellar modelling}
\label{section: chapter1}

Convection is omnipresent in stars at some point in their evolution. For stars like the Sun, convection occurs in a thick envelope and thus is the dominant mixing and energy transport mechanism at the photosphere. Very low-mass stars, such as M dwarfs, are fully convective throughout their interior. Whereas for compact stars, such as white dwarfs, conduction is an important energy transport mechanism. At the other end of the mass-scale in the upper part of the HR diagram, the hydrogen-burning cores of high-mass main-sequence stars are convective and their envelopes are radiative. During the main sequence the envelopes of massive stars are dominated by radiation, but convection becomes increasingly more important in their envelopes as they evolve off the main sequence. This means that for all stars in the HR~diagram the definition and numerical implementation of convection is important in calculating stellar structure models. However, convection is inherently a 3D process, but most state-of-the-art evolution models are 1D, with some being 2D. This means the optimum 1D representation and numerical prescription for convection is a topic of ongoing investigation.

In chapter~1 of this Special Issue, Joyce \& Tayar (2023) \citep{Joyce2023a} provide a detailed overview of convection inside stars. Specifically, they focus on the numerical prescription known as Mixing Length Theory (MLT; \citep{Bohm-Vitense1958}) commonly used to define convective regions within stellar structure models and to parameterise convection in 1D. MLT for convection is a formalism analogous to having a mean free path of a fluid parcel in thermodynamics, and specifies the distance such a parcel travels before fully mixing with its surrounding fluid. Yet the MLT formalism is only described in 1D which limits its application to 3D convective zones in stellar interiors (e.g. \citep{Arnett2015c}). Moreover, the acceleration of convective fluid parcels do not instantaneously become zero at the boundary of convective and radiative regions, meaning that MLT is insufficient to parametrise mixing in 1D at such boundaries.

There has been tremendous effort in calibrating MLT using observations, which has been hugely successful using helioseismology -- the study of the Sun's interior from its resonant pulsation frequencies \citep{C-D2002, ASTERO_BOOK, Basu2016}. The study of the interior physics using pulsations in stars other than the Sun is called asteroseismology \citep{ASTERO_BOOK, Kurtz2022a}. This field of stellar astrophysics has greatly expanded in recent years thanks to high-precision space photometry, such as from the Kepler mission \citep{Borucki2010, Koch2010}. It has provided tight constraints on the masses, radii, and ages of thousands of stars \citep{Chaplin2013c, Hekker2017a, SilvaAguirre2017a, Garcia_R_2019, Bowman2020c}, but also calibrated their interior rotation and mixing profiles, and angular momentum transport mechanisms \citep{Eggenberger2012c, Eggenberger2017a, Eggenberger2019a, Eggenberger2019c, Li_G_2020a, Aerts2021a, Pedersen2021a, Pedersen2022b, Mombarg2023a*}.

Asteroseismology combined with additional observables from spectroscopy and interferometry have also been able to provide constraints on MLT \citep{Tayar2017, Joyce2018a, Joyce2018b}. However, this remains a challenging endeavour that can only be achieved as of today for the brightest low-mass stars. A direct observational calibration of MLT for interior convection is currently beyond reach, but asteroseismology is able to provide constraints on this important physical ingredient of stellar structure. For example, asteroseismology can provide an empirical constraint on the mass and radius of convective cores in intermediate-mass \citep{Moravveji2015b, Moravveji2016b, Deheuvels2016a, Bellinger2019c, Mombarg2019a, Mombarg2020a, Mombarg2021a, Michielsen2021a, Pedersen2022a} and high-mass stars \citep{Dupret2004b, Mazumdar2006d, Briquet2007e, Briquet2012, Salmon2022a, Salmon2022b, Burssens2023a}.


\section{Chapter 2: Opacities and atomic diffusion}
\label{section: chapter2}

When radiation passes through a gas, photons are removed by scattering and absorption with the efficiency coefficient of these processes termed opacity. In reality, a star's opacity is a function of its chemical composition and the thermodynamic state of its constituent gas. Thus opacity influences how stars are formed in the earliest phases of their lives, because of the interaction between photons and atoms, but also how stars evolve and end their lives. For example, diffusion processes, such as radiative levitation, depend on stellar opacity mix and transport chemical species within stellar interiors. Therefore such processes are important contributors to uncertainties when inferring stellar ages and chemical abundances (see e.g. \cite{Dotter2017, Semenova2020a}). 

Opacity data are an important ingredient to essentially all forms of stellar structure modelling and all stars across the HR~diagram, and remain a primary contributor to the uncertainty of the heavy element content of our own Sun \cite{Asplund2009}. Stellar evolution models typically include and thus implicitly depend on an opacity table with a chosen number of specific chemical elements as main contributors, and a fixed chemical mixture for heavy elements (e.g. C, N and O). The construction of precise and accurate opacity tables for use within stellar physics is a herculean task within the atomic physics community \citep{Iglesias1993, Iglesias1996, Seaton1994, Seaton2005}, with which stellar astrophysics has great synergy.

Nuclear fusion, chemical mixing and atomic diffusion processes redistribute sources of opacity within a star, with the different processes taking place on different time scales. Since a full calculation of stellar opacity taking such processes into account is computationally expensive, especially for each time-step of a stellar evolution model, a common simplification is to calculate a mean opacity averaged across all wavelengths. A commonly used approach is to calculate the Rosseland mean opacity, which is a harmonic weighted average over all chemical species. Moreover, the full implementation of all atomic diffusion processes, which are inherently chemical-element specific, in evolutionary models can be computationally demanding, so not all of them are always included because of computation time arguments. Such a numerical simplification has validity that depends on the type of star being studied. 

For example, the Rosseland mean opacities can no longer be employed in the outermost layers of the most massive stars since the effective bandwidth of spectral lines are broadened by an accelerating velocity field from their line-driven winds. In this regime, the flux-weighted mean opacity including all relevant ions needs to be considered (see e.g. \cite{Sander2020a, Poniatowski2022a}). Recent work has demonstrated the divergence in applicability of Rosseland mean opacities and conclude that to use flux-weighted mean opacities correctly then hydrodynamical simulations in non-local thermodynamic equilibrium (NLTE) are needed \citep{Sander2020a}.

On the other hand, for accurate opacity profiles, spectroscopic surface abundances and ages (etc) of stars across the HR~diagram, all diffusion (and mixing) mechanisms should be considered in structure and evolution models when comparing to observations. Moreover, an investigation of different opacity tables and their impact on stellar structure is a valuable exercise. For example, stellar structure differences caused by different opacity tables with and without diffusion processes is known to significantly impact the location of pulsation instability regions in the HR~diagram \cite{Pamyat1999b, Walczak2015, Paxton2015, Moravveji2016a}, which can be probed using asteroseismology (see e.g. \cite{Turcotte2000c, Mombarg2022a}). 

In chapter 2 of this Special Issue, Alecian \& Deal (2023) \citep{Alecian_G_2023b} provide a detailed discussion of opacities and atomic diffusion inside stars. This includes a comparison of the different opacity tables available in the literature available for stellar modelling, their domains of applicability, and impact on opacity profiles in the context of providing accurate stellar structure and evolution models.


\section{Chapter 3: Magnetism in high-mass stars}
\label{section: chapter3}

Despite its importance and impact on stellar structure, magnetic fields are an important, yet typically missing, ingredient in evolution models. Observational constraints on the strength and geometry of magnetic fields, both at the surface and in the deep interiors, are generally lacking for the vast majority of stars. 

For low-mass stars like the Sun, their large convective envelopes during the main sequence produce weak global magnetic fields, but strong local fields in the form of sunspots through a dynamo process. The reader is referred to literature reviews of stellar magnetism in low-mass stars \citep{Borra1982a, Donati2009, Kochukhov2021d}. Although not explicitly covered in depth in this Special Issue, the consequences of magnetic fields, and their interplay with rotation and convection in stellar envelopes and exoplanetary systems are critical for our understanding of low-mass stars. For example, processes such as mixing and angular momentum evolution depend in part on the presence of a magnetic field \citep{Aerts2019b}. Moreover, the analysis of co-rotating magnetic spots at the photospheres of low-mass stars leads to detectable signatures of rotational modulation, which allows the rotation period of such stars to be measured. In turn, the co-evolution of rotation and magnetism as a function of time gave rise to gyrochronology --- the idea that stellar ages can be inferred from rotation periods --- first recognized as the Skumanich law \citep{Skumanich1972} and later refined into a tool \citep{Barnes2007}. On the other hand, pulsations in low-mass stars allow constraints on the strength and geometry of interior magnetic fields \citep{Fuller2015, Bugnet2021a, Bugnet2022c, Li_G_2022a}. The observational direct or indirect inference of magnetic fields in evolved low-mass stars, such as red giants, is also an exciting and novel prospect.

On the other hand, the origin and consequences of magnetic fields in massive stars are not as well understood compared to low-mass stars. In the last couple of decades, there has been a large effort and major progress in detecting surface magnetic fields in early-type stars (i.e. spectral types O and B). Ground-based surveys using spectropolarimetry such as the Magnetism in Massive Stars (MiMeS) \cite{Wade2016a} and Binarity and Magnetic Interactions in various classes of Stars (BinaMIcS) \cite{Alecian_E_2015} consortia established that large-scale magnetic fields with strengths between approximately tens of Gauss and tens of kG, which are predominantly dipolar in topology, exist at the surface of about 10\% of massive main-sequence stars. However, the origin of such strong and globally organised magnetic fields, whether they are of fossil origin (e.g. \cite{Borra1982a}) or are a result of stellar mergers (e.g. \cite{Schneider_F_2019a}), is yet to be generally established. Nor is it known how such fields evolve throughout a star's lifetime, because a single surface measurement defines only one epoch during a star's evolution. What is clear, however, is that the presence of a (strong) magnetic field inside a massive star has a significant impact on its interior rotation and thus mixing profiles compared to a non-magnetic star. Thus the evolution of magnetic massive stars strongly differ to non-magnetic stars (see \citep{Keszthelyi2019, Keszthelyi2020a, Keszthelyi2021a}).

Of course, almost all constraints of magnetic fields for massive stars are limited to their surface properties (e.g. strength and geometry) when using spectropolarimetry. The most promising method to diagnose the interior magnetic field properties and their impact on stellar evolution arise from magneto-asteroseismology --- the study of pulsations and their interaction with magnetic fields. First applications of this technique suggest that magnetic and non-magnetic massive stars have different interior mixing properties, with evidence for the presence of a magnetic field suppressing mixing in the near-core region of massive stars \citep{Briquet2012, Buysschaert2018c}. Most importantly, a novel and recent proof-of-concept study combining magneto-asteroseismology with numerical magnetohydrodynamical simulations delivered the first inference of the near-core magnetic field strength in the main-sequence early-type star HD~43317 \cite{Lecoanet2022a}, which is the currently the only confirmed strongly magnetic early-type star pulsating in gravity modes to have undergone asteroseismic modelling \citep{Papics2012a, Briquet2013, Buysschaert2017b, Buysschaert2018c}.

In the current era of large-scale time-series photometric surveys, such as Kepler \citep{Borucki2010, Koch2010}, TESS \citep{Ricker2015} and soon also PLATO \citep{Rauer2014} delivering years-long high-precision light curves and pulsation frequencies for tens of thousands of stars, the break-through application of magneto-asteroseismology to many more massive stars has just begun. The recent advances in theoretical, numerical and observational work focusing on magnetic fields in massive stars is described in Chapter~3 of this Special Issue by Keszthelyi (2023) \citep{Keszthelyi2023a}.


\section{Chapter 4: Multi-D simulations of core convection}
\label{section: chapter4}

In addition to traditional observational techniques and comparison to evolution models, complementary synthetic observables predicted and inferred from numerical simulations are also extremely valuable when studying the physics of stellar structure. 

The long time scales associated with nuclear burning processes prohibit numerical simulations of convection covering the duration of stellar evolution. However, simulations studying dynamical processes comparable to hydrostatic timescales are much more feasible. Therefore, simulations of relatively rapid processes such as convection and pulsations, which are perturbations to the equilibrium structure, yield prescriptions for stellar structure that can be implemented into 1D stellar evolution models. Moreover, evolution models necessarily take time steps that are orders of magnitude larger than the typical time scales of convection in most stars, and thus seek to include only the time-averaged net effects of dynamical processes such as convection, mixing and wave generation. Numerical simulations are a unique method of testing the validity of assumptions in stellar evolution models, and provide improved prescriptions where needed (see e.g. \cite{Freytag1996b, Herwig2000c, Scott2021a, Herwig2023a*}).

Owing to the importance of convection, and in particular core convection in main-sequence massive stars, there has been significant progress in using multi-dimensional numerical simulations to study the impact of core convection in the last several years. The complexity of large-scale multi-dimensional numerical simulations of stellar interiors cannot be understated, and owing to choices and subtleties of the different numerical codes available, direct one-to-one comparisons of results are not always possible. Recently, the first detailed comparison of several numerical codes to solve the same hydrodynamical problem was published \citep{Andrassy2022a}. This is a laudable effort and emphasises the need to perform such consistency checks when running and comparing codes, and when comparing synthetic observables to observations.

In chapter~4 of this Special Issue, Lecoanet \& Edelmann (2023) \citep{Lecoanet2023a} explore the literature of multidimensional simulations of core convection in main-sequence stars. In addition to providing a firm mathematical grounding in the physics, they also discuss and elucidate the differences and similarities in various numerical setups and their implications for the resultant synthetic observables. Given their importance for mixing and angular momentum transport within stellar interiors, the generation and propagation of waves excited by turbulent convection in the core of massive stars is discussed in detail. Moreover, the importance of combining perspectives from the observational and hydrodynamical communities is perhaps best evidenced by the recent interest in understanding the physical origin of the ubiquitous stochastic low-frequency (SLF) variability in massive stars \citep{Bowman2019a, Bowman2019b, Bowman2020b, Bowman2022e}. For example, numerical simulations and theoretical studies have proposed gravity waves excited by the convective core, and/or the dynamics of massive star envelopes and atmospheres as plausible mechanisms to explain SLF variability (see e.g. \cite{Rogers2013b, Rogers2015, Rogers2017c, Krticka2018e, Krticka2021b, Edelmann2019a, Cantiello2021b, Schultz2022a, Herwig2023a*, Thompson_W_2023a*}).


\section{Chapter 5: Convective boundary mixing in main-sequence stars}
\label{section: chapter5}

In addition to an incomplete theory of convection and how to implement it in 1D evolutionary models, the specific issue of the amount and shape of the mixing profile at the boundary of convective and radiative zones remains a primary unresolved uncertainty in stellar structure theory (e.g. \citep{Kaiser2020a, Scott2021a}). This problem, known as convective boundary mixing (CBM), cannot be derived from first principles and encompasses several possible physical scenarios. One mechanism is convective penetration in which the dynamics of convective motions significantly extend into an overlying radiative zone causing an extended convection zone \citep{Zahn1991, Augustson2019a, Anders_E_2022a, Jermyn2022e}. Additionally, mixing at the convective-radiative boundary can give rise to increased chemical entrainment, which alters the chemical composition in the zone just beyond the convective boundary in the radiative zone.

There are two main techniques that have made significant strides in our understanding of CBM: (i) hydrodynamical simulations and (ii) observations. For the former, the ability to vary parameters within different numerical setups allows one to study the fundamental physics of convection, and moreover the physics of the boundary layer of convective and radiative regions. This then allows physical prescriptions from multi-dimensional numerical simulations to be implemented in 1D evolution models (see e.g. \cite{Scott2021a, Anders_E_2022a}).

In terms of observational constraints, the comparison of various observables to theoretical predictions from (evolution) models reveal discrepancies in the physics of convection and thus also CBM. For example, spectroscopy, binary modelling and asteroseismology have all independently highlighted not only the importance of CBM in stellar structure and evolution, but also have been able to quantify, to differing levels of precision, the shape of the mixing profile of the CBM region \citep{Claret2016b, Martinet2021a, Johnston2021b}. In particular, gravity-mode asteroseismology offers a route to precise calibration of the size and shape of CBM owing to the unique probing power of such pulsations to the convective-radiative boundary region in main-sequence early-type stars (e.g. \citep{Moravveji2015b, Mombarg2019a, Pedersen2021a}). In Chapter 5, Anders \& Pedersen (2023) \citep{Anders_E_2023a} discuss the recent advances in combining knowledge from state-of-the-art hydrodynamical simulations and observations of CBM, with a particular focus for main-sequence stars as these are the most common in the Universe.


\section{Chapter 6: Radiation dominated envelopes of massive stars}
\label{section: chapter6}

While convection is relevant in most stars, the uncertainties in energy transport are nowhere as large as for the outer layers of the most massive stars. In standard 1D stellar evolution models, stars close to the Eddington limit, which is the maximum luminosity a star can have such that outward radiation balances the force of gravity acting inward, stellar models develop rather peculiar structures.

For example, an additional source of opacity from, for example, iron group elements can make massive stars exceed the Eddington limit in their outer layers, making convection inefficient and leads to hugely inflated envelopes with density inversions (see Fig.\,1 from \cite{Graefener2012}) in these super-adiabatic layers.  Perhaps efficient convection can solve this particular super-adiabaticity problem. For instance, in the 1D stellar evolution code MESA (i.e. Modules for Experiments in Stellar Astrophysics; \citep{Paxton2011, Paxton2013, Paxton2015, Paxton2018, Paxton2019, Jermyn2023a}) efficient convection is artificially implemented using the so-called MLT$++$ functionality. It is as yet uncertain what happens in Nature. However, alternative solutions involve the creation of porous structures \citep{Shaviv2000, Cantiello2009, Jiang2015}, or the launch of a radiation-driven wind for stars close to the Eddington limit \citep{Vink2011, Graefener2017, Moens2022}. 

In Chapter 6 of this Special Issue, Jiang (2023) \citep{Jiang_Y_2023a**} discusses the state-of-the art of 3D radiation-dominated envelopes. In reality, the envelopes and stellar winds probably interact with one another, either by winds removing the inflated layer \citep{Petrovic2006}, or by other wind-envelope interactions, which may play a role in the S\,Doradus variability of Luminous Blue Variables \citep{Grassitelli2021}. Future 3D envelope and wind simulations are required to solve the envelope inflation question. The implications are huge, because inflated stars are thought to be luminous but cool, whereas a reduction in the influence of inflation would make very massive stars luminous and hot, thereby producing copious amounts of ionising radiation.


\section{Discussion and perspectives}
\label{section: discussion}

In our introductory chapter of this Special Issue on `The Structure and Evolution of Stars', we have provided a flavour of the important aspects of each subsequent chapter. Of course, this Special Issue does not provide an exhaustive discussion of all important aspects of stellar structure theory, but does provide an overview of the topics that have undergone major progress in the last decade and for which significant advancement is expected in the years to come.

A major factor driving improvement has been the recent availability of pristine time-series photometric data from space missions, such as Kepler and TESS. These data have allowed ultra-precise measurements of stellar interiors to be made and confronted with state-of-the-art models. For example, inference of interior rotation profiles \citep{Aerts2019b}, the size of convective cores \citep{Mombarg2019a, Burssens2023a}, efficiency of various mixing processes \citep{Pedersen2021a}, and even magnetic field strength and geometry \citep{Buysschaert2018c, Lecoanet2022a, Li_G_2022a} are now within reach for a large number of stars across the HR~diagram thanks to asteroseismology \citep{Aerts2021a}. With the anticipation of the approval of a third extended mission for TESS and the upcoming PLATO mission, we expect asteroseismology to go from strength to strength in the next several years.

Moreover, it is the unique synergy of observations, numerical simulations and evolutionary modelling that gives rise to the tightest constraints on stellar interiors, and thus an improvement in our understanding of the physical processes truly at work. In preparing this Special Issue, we have paid particular attention in inviting researchers whose expertise lie at the intersection of these three circles within the Venn diagram of stellar astrophysics in the hopes to help bridge these valuable communities.

\vspace{6pt} 




\authorcontributions{All authors are guest editors of the Special Issue entitled `The Structure and Evolution of Stars' and contributed to the writing of this chapter, as well as editing the other chapters in this Special Issue as guest editors.}

\funding{DMB gratefully acknowledges funding from the Research Foundation Flanders (FWO) by means of a senior postdoctoral fellowship (grant agreement No. 1286521N), and an FWO long stay travel grant (agreement No. V411621N) that supported his stay at the Kavli Institute for Theoretical Physics (KITP), University of California Santa Barbara (UCSB), USA between October and December 2021. JSV gratefully acknowledges
support from STFC via grant ST/V000233/1. This work was supported in part by the National Science Foundation (NSF) under Grant Number NSF PHY-1748958.}

\dataavailability{Not applicable.}  

\acknowledgments{The authors are grateful to the staff and scientists at the Kavli Institute for Theoretical Physics (KITP), University of California Santa Barbara (UCSB), USA, and to the organisers of the ‘Probes of Transport in Stars’ program between October and December 2021, at which the first conception of this Special Issue was synthesised.}

\conflictsofinterest{The authors declare no conflict of interest.}  



\abbreviations{Abbreviations}{
The following abbreviations are used in this manuscript:\\
\noindent 
\begin{tabular}{@{}ll}
BinaMIcS & Binarity and Magnetic Interactions in various classes of Stars \\
CBM & Convective Boundary Mixing \\
HR & Hertzsprung--Russell \\
NLTE & Non-Local Thermodynamic Equilibrium \\
MiMeS & Magnetism in Massive Stars \\
MLT & Mixing Length Theory \\
SLF & Stochastic Low Frequency \\
\end{tabular}
}

\begin{adjustwidth}{-\extralength}{0cm}

\reftitle{References}


\bibliography{bibliography.bib}

%


\end{adjustwidth}
\end{document}